\documentclass[twocolumn,pre,showpacs]{revtex4-2}
\usepackage{graphicx}
\usepackage{amsmath}
\usepackage{amsfonts}
\usepackage{mathtools}
\usepackage{subfigure}

\begin{document}
\title{Cram\'er–Rao Inequality Generalizes the Equilibrium Energy Fluctuation–Response Relation to Nonequilibrium Steady States}

\author{Shaofan Liu$^1$, Raphael Chetrite$^{2}$, and Andre C. Barato$^1$}
\affiliation{$^1$ Department of Physics, University of Houston, Houston, Texas 77204, USA\\
$^2$ CNRS Laboratoire Ypatia des Sciences Mathématiques (LYSM), Rome, Italy
}

\parskip 1mm
\def\d{{\rm d}}
\def\Ps{{P_{\scriptscriptstyle \hspace{-0.3mm} s}}}
\def\MF{{\mbox{\tiny \rm \hspace{-0.3mm} MF}}}
\def\ts{\tau_{\textrm{sig}}}
\def\tos{\tau_{\textrm{osc}}}
\def\Ib{\mathcal{I}_\beta}
\def\FlucE{\textrm{Var}(E)}
\def\RE{\partial_\beta\langle E\rangle}

\begin{abstract}

 Equilibrium steady states are fully characterized by the Boltzmann distribution, which depends only on the energy and the temperature. In contrast,
 nonequilibrium steady states, which describe a wide range of physical and biological phenomena, are generically much more complex, 
 their stationary distributions can depend on a zoo of kinetic parameters. A central challenge of nonequilibrium statistical 
 mechanics is the identification of universal relations that hold for nonequilibrium steady states despite this complexity. 
 We show here that a form of the Cram\'er–Rao inequality applies to arbitrary nonequilibrium steady‑state distributions and contains no explicit dependence on kinetic parameters.
 A main original feature in our observation is that this inequality becomes a well-known fluctuation-response relation  in equilibrium.  
 The inequality involves three well‑defined quantities,  the fluctuations of the energy, the response given by the derivative of the average energy with respect to the inverse temperature, 
 and the Fisher information with respect to the inverse temperature. We illustrate this inequality using three representative models, an active particle in a harmonic potential,
 a simple scheme for kinetic proofreading, and a one-dimensional model for heat conduction between two baths.

\end{abstract}

\maketitle
\section{Introduction}

Equilibrium statistical mechanics \cite{reic09} is characterized by the key property that the probability distribution of the states of a system in contact with 
a heat bath is given by the Boltzmann distribution. This distribution has the remarkable feature that it depends only on the energy of the states and 
the temperature of the bath. As a consequence, it leads to several fundamental properties, including standard fluctuation–dissipation relations, 
and enables the description of equilibrium macroscopic states in terms of a small number of thermodynamic variables, despite the exponentially 
large number of microscopic configurations.

Most processes in nature occur out of equilibrium. This motivates the extension of equilibrium concepts to nonequilibrium systems, with 
stochastic thermodynamics \cite{seif12} providing a prominent framework. Nonequilibrium steady states (NESS) form an important class of 
such systems, they are characterized by non-vanishing stationary fluxes and a time-independent probability distribution. 
In contrast to the Boltzmann distribution, the stationary distribution of a NESS generically depends on kinetic parameters. 
Even for systems with a small number of states, this dependence can lead to highly nontrivial distributions.

Despite the absence of the simplifying structure of equilibrium, a number of universal results for nonequilibrium 
systems have been established. The most prominent examples are fluctuation relations \cite{evan93,gall95,jarz97,croo99,lebo99,seif05}, 
which express symmetries of the probability distribution of entropy production along stochastic trajectories. Another important class of 
results is given by the thermodynamic uncertainty relation \cite{bara15} and related inequalities \cite{ging16,piet16,bara16,pole16,garr17,piet17b,ito20,horo20,fala20,piet22,dieb23,ohga23,meie25}, 
which bound the precision of thermodynamic currents by entropy production. Importantly, these results typically involve dynamical quantities
 and require knowledge beyond the stationary distribution.

Another line of research concerns general relations between fluctuations and response in nonequilibrium systems. Various forms of nonequilibrium fluctuation–dissipation relations 
have been derived based on small perturbations of the stochastic matrix \cite{agar72,chet08,marc08,gome09,baie09,seif10b,chet11}. These relations connect response functions to time-dependent 
correlation functions of the unperturbed system. More recently,  response relations and bounds have been obtained for nonequilibrium systems. In particular, constraints 
on the linear response to perturbations of transition rates have been derived in terms of thermodynamic driving \cite{owen20}, with applications to biochemical systems \cite{mart23,owen23,chun23}. 
Fluctuation–response equalities and inequalities have also been reported \cite{ptas24,ptas25,ptas26,ptas26b,kwon25,liu25}, as well as approaches based on 
trajectory-level information geometry leading to Cramér–Rao-type bounds on response \cite{dech20,zhen25,zhen25b}. 
In addition, structural constraints on the steady-state response of currents in Markov jump processes have been identified \cite{haru24,asly24,asly24b,dalc25,bebo16}.

In this paper, we introduce a generic inequality for nonequilibrium steady states as a direct consequence of the generalized Cram\'er-Rao bound (CRB) \cite{kay93,cove06}. 
In contrast to the above approaches, our result does not rely on any assumptions about the underlying dynamics and is 
independent of the kinetic parameters. In equilibrium, it reduces to a well-known equality, while away from equilibrium 
it provides a universal constraint on the stationary distribution. While the application of the CRB here is straightforward, the novelty relies on 
the observation that  the CRB considered here becomes  a well-known fluctuation-reponse equality in equilibrium. 

The inequality relates three quantities: the variance of the energy, the Fisher information \cite{kay93,cove06} of 
the NESS distribution with respect to the inverse temperature, and the response given by the derivative of the average energy with respect to the inverse temperature. 
We illustrate our observation in three paradigmatic systems: an active run-and-tumble particle confined in a one-dimensional  harmonic potential, a kinetic proofreading network, and 
a one-dimensional interacting many-body model for heat conduction. The latter example demonstrates that the inequality remains valid in interacting many-body systems.

The paper is organized as follows. In Sec. \ref{sec2}, we show the version of the CRB we consider here and how it becomes the fluctuation-response relation in equilibrium. We also 
show that a linearized version of the inequality can be proved with an alternative method. In Sec. \ref{sec3}, we illustrate the inequality with exact calculations for three different models.
We conclude in Sec. \ref{sec4}. Detailed calculations for the models from \ref{sec3} are shown in App. \ref{appA}.

\section{Cram\'er-Rao inequality for  a NESS}
\label{sec2}

\subsection{Main result}

There are two main physical assumptions in our work. First, even out of equilibrium, 
the notion of temperature remains well defined. The dynamical rules governing the time evolution of the system 
must be consistent with this temperature. Within stochastic thermodynamics, where systems are described by 
Markovian dynamics, this requirement is encoded in the generalized detailed balance condition for the transition rates \cite{seif12}. 

Second, there exists a reference equilibrium distribution corresponding to the same temperature and to well-defined 
energies of the states. The system is driven out of equilibrium by nonzero thermodynamic affinities, when these affinities are set to zero, 
the stationary distribution reduces to the equilibrium Boltzmann distribution. These two assumptions are naturally satisfied within stochastic thermodynamics, and, therefore, 
our result applies to this framework for systems with either discrete or continuous states, and for both overdamped and underdamped dynamics.

We consider a generic NESS characterized by the distribution $P_i$ (of state $i$), with energy $E_i$, and the inverse temperature $\beta \equiv 1/(k_B T)$, 
where $k_B$ is Boltzmann's constant. We set $k_B=1$ throughout. For convenience, we consider discrete states but the same observations below apply for continuous states.
 The inequality analyzed here is obtained by taking  the energy as an observable and 
the inverse temperature $\beta$ as parameter that influences the distribution, the  generalized CRB \cite{kay93,cove06} for this choice can be written as  
\begin{equation}
\left|\partial_\beta \langle E \rangle\right|\le \sqrt{\mathrm{Var}(E) \mathcal{I}_\beta},
\label{eqmain}
\end{equation}
where 
\begin{equation}
\langle E \rangle \equiv \sum_i P_i E_i
\end{equation}
is the average energy, 
\begin{equation}
\mathrm{Var}(E) \equiv \langle E^2 \rangle - \langle E \rangle^2
\end{equation}
is the variance of the energy, and
\begin{equation}
\mathcal{I}_\beta \equiv \sum_i P_i \left(\partial_\beta \ln P_i\right)^2=-\langle \partial^2_\beta \ln P\rangle
\label{eqdefI}
\end{equation}
is the Fisher information of the distribution with respect to $\beta$. The generalized CRB in equation \eqref{eqmain} follows from 
the Cauchy-Schwarz inequality $\mathrm{Var}(E) \mathrm{Var}(\partial_\beta \ln P)\ge \langle (E-\langle E\rangle)(\partial_\beta \ln P-\langle\partial_\beta \ln P\rangle)\rangle^2$
and the fact that $\langle\partial_\beta \ln P\rangle=0$. The word ``generalized" here reflects the fact that the original CRB uses an unbiased estimator of $\beta$, with the response term equals to $1$ \cite{kay93,cove06}.

As our main observation, in equilibrium, this inequality is saturated and reduces to the standard energy fluctuation-response relation 
\begin{equation}
-\partial_\beta \langle E \rangle= \mathrm{Var}(E).
\label{eqFReq}
\end{equation}
Indeed, for the Boltzmann distribution $P_i \propto e^{-\beta E_i}$, the Fisher information satisfies $\mathcal{I}_\beta = \mathrm{Var}(E)$. We point out 
that in equilibrium $-\partial_\beta \langle E \rangle$ is positive. Therefore, the CRB in Eq. \eqref{eqmain} is a generalization of the equilibrium fluctuation-response 
relation in Eq. \eqref{eqFReq} that applies to generic nonequilibrium steady states.  To quantify how close a system is to these bounds in applications we define the ratio
\begin{equation}
\eta_1 \equiv \frac{\left|\partial_\beta \langle E \rangle\right|}{\sqrt{\mathrm{Var}(E)  \mathcal{I}_\beta}} \le 1.
\label{eqeta1}
\end{equation}

It is possible to linearize the inequality in Eq. \eqref{eqmain} by using the relation $\sqrt{\mathrm{Var}(E) \mathcal{I}_\beta}\le (\mathrm{Var}(E) +\mathcal{I}_\beta)/2$, which leads to
\begin{equation}
2\left|\partial_\beta \langle E \rangle\right|\le \mathrm{Var}(E) +\mathcal{I}_\beta.
\label{eqmainlin}
\end{equation}
For this inequality we use the ratio 
\begin{equation}
\eta_2 \equiv \frac{2\,\left|\partial_\beta \langle E \rangle\right|}{\mathrm{Var}(E) + \mathcal{I}_\beta}\le \eta_1 \le 1.
\label{eqeta2}
\end{equation}

Notably, the inequality in Eq. \eqref{eqmain} does not rely on any specific form of the underlying dynamics, such as particular transition 
rates or equations of motion. It only requires that the dynamics are consistent with an underlying equilibrium distribution of Boltzmann 
form at temperature $\beta^{-1}$, which is recovered when thermodynamic affinities vanish.
In contrast, recent results on nonequilibrium steady-state response \cite{haru24,asly24,asly24b,dalc25,bebo16} 
are formulated within the framework of Markov jump processes and involve perturbations of transition rates, 
leading to response relations that depend explicitly on kinetic parameters. We  also note that related works have 
introduced definitions of heat capacity in nonequilibrium steady states \cite{boks11,mand13,maes19,boge25}, 
where quantities such as the thermal response $\partial_\beta \langle E \rangle$ play a central role.

\subsection{Alternative proof of the linear inequality and  generalization to other thermodynamic parameters}

It is possible to prove the linear inequality in Eq. \eqref{eqmainlin} without using the CRB. 
The proof proceeds by introducing the variable $\phi_i \equiv \ln\left(\frac{P_i}{P_i^{\mathrm{eq}}}\right)$,
where $P_i$ denotes the nonequilibrium steady-state (NESS) distribution and $P_i^{\mathrm{eq}} \propto \mathrm{e}^{-\beta E_i}$ 
is the equilibrium Boltzmann distribution obtained by setting all thermodynamic affinities to zero. 
The first step of the proof is the identity
\begin{equation}
\mathrm{Var}(E) + \partial_\beta \langle E \rangle 
= \mathrm{Var}(\partial_\beta \phi) 
+ \langle \partial_\beta^2 \phi \rangle 
- \partial_\beta \langle \partial_\beta \phi \rangle,
\label{eqmainequality}
\end{equation}
which can be verified by direct evaluation of the right-hand side. Since $\mathrm{Var}(\partial_\beta \phi) \ge 0$, it follows that
\begin{equation}
\mathrm{Var}(E) + \partial_\beta \langle E \rangle 
\ge 
\langle \partial_\beta^2 \phi \rangle 
- \partial_\beta \langle \partial_\beta \phi \rangle.
\end{equation}
Using the relation $\langle \partial_\beta^2 \phi \rangle - \partial_\beta \langle \partial_\beta \phi \rangle= -\partial_\beta \langle E \rangle - \mathcal{I}_\beta$,
we obtain the inequality in Eq.~(\ref{eqmainlin}).

The inequality in Eq. \eqref{eqmain} can be generalized to other thermodynamic conjugate pairs. For example, instead of temperature and energy,
 one can consider the chemical potential $\mu$ and the particle number $N_i$ associated with state $i$. In equilibrium, 
 for a system exchanging particles with a reservoir at chemical potential $\mu$, the grand canonical distribution is
$P_i^{\mathrm{eq}} \propto \mathrm{e}^{-\beta E_i + \beta \mu N_i}$. If the system is driven out of equilibrium by thermodynamic affinities, the CRB for a NESS can be written as 
\begin{equation}
\left|\partial_\mu \langle N \rangle\right|\le \sqrt{\mathrm{Var}(N)  \mathcal{I}_\mu}. 
\label{eqmain2}
\end{equation}
At equilibrium, where $ \mathcal{I}_\mu= \beta^2 \mathrm{Var}(N)$ this inequality reduces to the known identity 
\begin{equation}
\partial_\mu \langle N \rangle= \beta \mathrm{Var}(N).
\end{equation}
Therefore, this idea that the CRB becomes a fluctuation-respsonse relation in equilibrium, naturally extends to any pair consisting of an extensive 
observable and its conjugate control parameter appearing in the equilibrium distribution.

\section{Case studies}
\label{sec3}

In this section, we calculate the three quantities in the CRB and the ratios $\eta_1$ quantifying the saturation 
of the CRB and $\eta_2$ quantifying the linearized version of the inequality. These calculations are performed for 
models of NESS that allows for exact results. All these quantities are plotted as a function of the 
thermodynamic affinity for the particular model. If this affinity is zero, the system is in equilibrium and 
the inequality is saturated.

\subsection{Active Particle in a harmonic potential}

Our first example is a one-dimensional active run-and-tumble particle confined in a harmonic potential in the presence of thermal noise \cite{garc21} (see also \cite{dahr19} for exact results in the absence of thermal noise). 
We include thermal noise to ensure the existence of a well-defined equilibrium reference state in the absence of activity. The particle switches between two propulsion states with forces $\pm f$ at a rate $\alpha$. 
The parameter $f$ plays the role of a thermodynamic affinity: for $f=0$, the system relaxes to an equilibrium stationary state, while for $f \neq 0$, it reaches a NESS.

The dynamics of the probability densities $P_\pm(x,t)$, where the subscript refers to the sign of the force, are governed by the coupled Fokker–Planck equations
\begin{equation}
\partial_t P_\pm(x,t)
= 
- \partial_x J_\pm(x,t)
- \alpha P_\pm(x,t)
+ \alpha P_\mp(x,t),
\label{eqmasterAP}
\end{equation}
where the probability currents are
\begin{equation}
J_\pm(x,t) \equiv \gamma^{-1} \big(\pm f + F(x)\big)\, P_\pm(x,t)-(\beta\gamma)^{-1}\frac{d}{dx}P_{\pm}(x,t),
\end{equation}
and the restoring force is $F(x) = -\kappa x$. Here, $\gamma^{-1}$ is the particle mobility and $\kappa$ is the stiffness of the harmonic potential.

This model can be solved exactly, allowing for analytical expressions for several quantities entering Eq.~\eqref{eqmain}. In particular, the thermal response $\partial_\beta \langle E \rangle$ 
is independent of the driving force $f$. The Fisher information $\mathcal{I}_\beta$ can be expressed as a one-dimensional integral, which we evaluate numerically. 
Details of these calculations are provided in App. \ref{appA}.


\begin{figure}[t]
  \centering
  \includegraphics[width=\columnwidth]{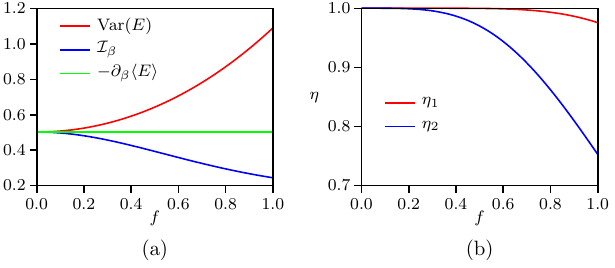}
  \caption{
  Plots for the active particle in a harmonic potential. (a) Quantities entering Eq.~\eqref{eqmain} as a function of the thermodynamic affinity $f$. (b) Ratios $\eta_1$ and $\eta_2$ as a functions of the thermodynamic affinity $f$.
  }
  \label{Fig1}
\end{figure}

In Fig.~\ref{Fig1}, we plot the three quantities entering Eq.~\eqref{eqmain} as a function of the driving $f$. The response $\partial_\beta \langle E \rangle$ is independent of $f$, as shown in Eq. \eqref{reponseharmonic}, while the fluctuations $\mathrm{Var}(E)$ increase with $f$, and the Fisher information $\mathcal{I}_\beta$ decreases with increasing affinity. The enhancement of fluctuations relative to the suppression of Fisher information is such that the inequality in Eq.~\eqref{eqmain} is satisfied. The deviation from saturation is quantified by the ratios $\eta_1$ and $\eta_2$. The bound is saturated in equilibrium ($f=0$), and the system moves further away from saturation as the driving $f$ increases. As we show next, this behavior is not universal and depends on the specific model under consideration.

\subsection{Model for kinetic proofreading}

We now consider a simple network for kinetic proofreading \cite{hopf74,nini75}. Kinetic proofreading is a nonequilibrium mechanism that reduces errors in the copying of biological information. We adopt the parametrization of the transition rates from Ref.~\cite{hart15}. The full set of rates and the method used to compute the NESS distribution exactly are described in App. \ref{appA}. Here, we summarize the model and present results illustrating our bound.

The system consists of an enzyme $\mathrm{E}$ and two substrates: a right one, denoted by $R$, and a wrong one, denoted by $W$. The model has five states: 
the free enzyme $\mathrm{E}$; the bound states $\mathrm{ER}$ and $\mathrm{EW}$; and the phosphorylated states $\mathrm{ER}^*$ and $\mathrm{EW}^*$. The energies of the states $\mathrm{E}$, $\mathrm{ER}$, 
and $\mathrm{ER}^*$ are set to zero, while the states $\mathrm{EW}$ and $\mathrm{EW}^*$ have energy $\Delta E$. We assume that discrimination between right and wrong substrates is purely energetic, with no entropic contribution. In more general kinetic proofreading schemes, the distinction is determined by a free-energy difference that may include entropic terms. Our inequality remains valid in that case, however, when computing the average energy and its fluctuations, one must use only the energetic contribution (excluding entropic terms).

The key feature of kinetic proofreading is the reduction of the copying error, defined as the ratio $P_{\mathrm{EW}^*}/P_{\mathrm{ER}^*}$. 
At equilibrium, this ratio is fixed by detailed balance and given by
$P_{\mathrm{EW}^*}/P_{\mathrm{ER}^*} = \mathrm{e}^{-\beta \Delta E}$. Out of equilibrium, the stationary distribution is modified in such a way 
that the error can be reduced below this equilibrium value. The nonequilibrium driving is provided by the free energy of ATP hydrolysis, denoted by $\Delta\mu$, 
which acts as the thermodynamic affinity. For $\Delta\mu=0$, the system is at equilibrium, whereas for $\Delta\mu \neq 0$ it reaches a NESS. 
The explicit expressions for the transition rates, which also depend on additional kinetic parameters, are given in App. \ref{appA}.

\begin{figure}[t]
  \centering
  \includegraphics[width=\columnwidth]{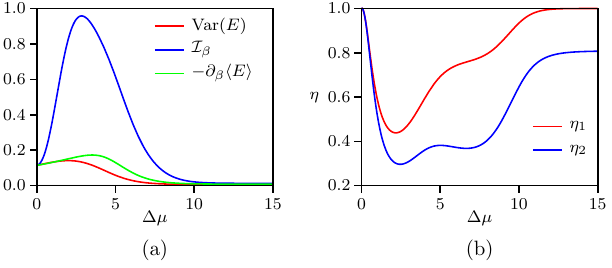}
  \caption{
  Plots for the kinetic proofreading model.
  (a) Quantities entering Eq.~(\ref{eqmain}) as a function of the thermodynamic affinity $\Delta\mu$.
  (b) Ratios $\eta_1$ and $\eta_2$  as functions of $\Delta\mu$.
  }
  \label{Fig2}
\end{figure}

In Fig.~\ref{Fig2}, we plot the three quantities appearing in Eq.~(\ref{eqmain}) as a function of the thermodynamic affinity $\Delta\mu$. 
For this model, the Fisher information $\mathcal{I}_\beta$ initially increases with $\Delta\mu$, reaches a maximum, and then decreases. 
The energy fluctuations $\mathrm{Var}(E)$ can become smaller than the response $-\partial_\beta \langle E \rangle$, nevertheless, 
our main inequality is always satisfied. The ratios $\eta_1$ and $\eta_2$ exhibit non-monotonic behavior, indicating that increasing the affinity $\Delta\mu$ 
does not necessarily drive the system further away from the bound. Thus, the quantities characterizing fluctuations, information, and response 
can display qualitatively different dependences on the driving for different models. In all cases, however, our main result remains valid.
Our examples thus far concern small systems. We now turn to a many-body system to further illustrate the generality of Eq.~(\ref{eqmain}).

\subsection{Kipnis–Marchioro–Presutti model}

We now consider the application of our result to a many-body system, namely the Kipnis--Marchioro--Presutti (KMP) 
model for heat conduction in one dimension \cite{kipn82,bert05,bert15b}. 
The system consists of $N$ sites, where the stochastic variable $e_i$ denotes the energy of site $i$. The left boundary 
site is in contact with a heat bath at temperature $T_-$, while the right boundary site is in contact with a heat bath at temperature $T_+$.
The bulk dynamics consists of stochastic energy exchange between neighboring sites: each pair of neighboring sites
 redistributes its total energy randomly, while conserving the sum of their energies. At the boundaries, the energy of 
 the boundary sites is refreshed according to the equilibrium distribution corresponding to the respective bath temperatures. 
 The full rules for the model, together with the exact calculation of the relevant quantities, are provided in App. \ref{appA}. The KMP model is a genuine many-body system, in which nonequilibrium conditions give rise to nonlocal correlations. These effects play a central role in the calculation of $\FlucE$ and $\Ib$. For our analysis, we employ macroscopic fluctuation theory \cite{bert05,bert15b}. To the best of our knowledge, the explicit calculation of $\Ib$ for this model, presented in App. \ref{appA}, is an original contribution.
 
\begin{figure}[t]
  \centering
  \includegraphics[width=\columnwidth]{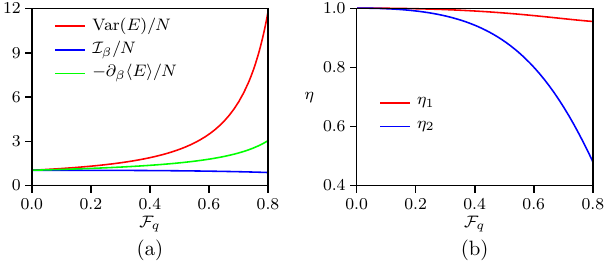}
  \caption{
  Plots for the KMP model.
  (a) Quantities entering Eq.~(\ref{eqmain}), divided by the system size $N$, as a function of the thermodynamic affinity $F_q$.
  (b) Ratios $\eta_1$ and $\eta_2$  as a function of $F_q$.
  }
  \label{Fig3}
\end{figure}

The results shown in Fig.~\ref{Fig3} are expressed in terms of the thermodynamic affinity $F_q = 1 - \frac{T_-}{T_+}$.
Without loss of generality, we consider the regime $0 \le F_q \le 1$, corresponding to $T_- \le T_+$. The case $F_q=0$ corresponds to thermal equilibrium ($T_- = T_+$), 
while the limit $F_q \to 1$ corresponds to $T_+ \to \infty$. We focus on the limit of large system size $N$ and plot the three quantities appearing in Eq.~(\ref{eqmain}), which are extensive, normalized by $N$.
For this model, the energy fluctuations are larger than the response $-\partial_\beta \langle E \rangle$, while the Fisher information $\mathcal{I}_\beta$ is smaller. 
The ratios $\eta_1$ and $\eta_2$ are a decreasing function of $F_q$. In the limit $F_q \to 1$, the energy fluctuations diverge as $(1 - F_q)^{-2}$, as shown in Eq. \eqref{eqFlucEKMP}.

In summary, the behavior of fluctuations, information, and response as functions of the thermodynamic driving differs across the three models considered. 
Two correspond to small systems, while the KMP model represents a genuine many-body system. Despite these differences, our inequality provides 
a unifying principle that holds in all cases.

\section{Conclusion}
\label{sec4}

We have identified the following property of the CRB for a NESS distribution.
It is an inequality  expressed in terms of physically meaningful observables associated with energy and temperature, and, 
notably, it does not involve kinetic parameters, despite the generally complex dependence of the NESS distribution on 
such parameters. The form of the CRB introduced here has the appealing property of reducing, at equilibrium, to the well-known equality 
relating energy fluctuations to the response of the average energy with respect to temperature. Hence, the CRB used here is a generalization 
of the equilibrium fluctuation-response relation to nonequilibrium steady states.

The inequality in Eq.~(\ref{eqmain}) can be interpreted as a bound on any of the three quantities appearing in it, depending
 on the context. It provides an upper bound on the thermal response of the average energy, a lower bound on energy fluctuations, 
 and a lower bound on the Fisher information, for instance, in problems involving the inference of temperature from a NESS distribution. 
 The generality of our version of the CRB is illustrated by its application to three different models, including both small systems and a 
 genuinely many-body system. Furthermore, the inequality is not restricted to energy and temperature: it applies more broadly to any 
 pair consisting of a control parameter and its conjugate extensive observable appearing in the equilibrium distribution, 
 such as chemical potential and particle number.

Several directions for future work emerge naturally. Applications of the inequality to specific models may provide further insight into 
their nonequilibrium properties. Extensions beyond classical stochastic thermodynamics, for example, to quantum open systems, 
appear plausible given the simplicity and generality of the proof. Finally, clarifying the relation between our result and recent advances
in nonequilibrium response theory constitutes an interesting direction for further investigation.

\appendix

\section{Calculations for the case studies}
\label{appA}

\subsection{Calculations for the active particle in a harmonic potential}

The master equation for this model in Eq.~\eqref{eqmasterAP} can be rewritten in terms of 
$P\equiv P_+ + P_-$ and $Q\equiv P_+ - P_-$. In the steady state, where time derivatives vanish, 
it reduces to a third-order ordinary differential equation for $P$, given by
\begin{align}
& \beta^{-2} \partial_x^3 P + 2 \beta^{-1} \kappa x \partial_x^2 P 
+ 2 \kappa x(\kappa - \alpha \gamma) P \nonumber \\
& \quad + \left(\kappa^2 x^2 + 3 \beta^{-1} \kappa - f^2 - 2 \beta^{-1} \alpha \gamma\right) \partial_x P = 0.
\end{align}
This equation can be solved by taking the Fourier transform of $P$, which yields the convolution form
\begin{equation}
P(x) = \frac{1}{\mathcal{Z}} \int_{-f/\kappa}^{f/\kappa} \mathrm{d}y \,
\mathrm{e}^{-\beta \kappa (x-y)^2 / 2}
\left[\left(\frac{f}{\kappa}\right)^2 - y^2 \right]^{\alpha \gamma / \kappa - 1},
\label{Px}
\end{equation}
where $\mathcal{Z} = \left(\frac{\beta \kappa}{2\pi}\right)^{-1/2}
\left(\frac{f}{\kappa}\right)^{2\alpha\gamma/\kappa-1}
\mathrm{B}\left(\frac{1}{2}, \frac{\alpha \gamma}{\kappa}\right),
$ and $\mathrm{B}(z_1, z_2)$ denotes the Beta function. The distribution $P(x)$ in the limit $f\to 0$, which corresponds to the equilibrium case, becomes $P(x)\propto \mathrm{e}^{-\beta \kappa x^2 / 2}$.

Since $E(x) = \kappa x^2/2$, the mean $\langle E \rangle$ and variance $\mathrm{Var}(E)$ can be obtained from the moments $\langle x^2 \rangle$ and $\langle x^4 \rangle$. 
To compute these moments, we exploit the convolution structure of the stationary distribution in Eq.~\eqref{Px}. The integrand in Eq.~\eqref{Px} can be interpreted as a conditional distribution of $x$ given $y$, with Gaussian weight proportional to $\mathrm{e}^{-\beta \kappa (x-y)^2/2}$, multiplied by a marginal distribution of $y$ proportional to $\left[\left(\frac{f}{\kappa}\right)^2 - y^2\right]^{\alpha \gamma/\kappa - 1}$.
As a result, moments of $x$ can be obtained by combining Gaussian moments with the moments of $y$. The latter can be evaluated in closed form using Beta-function integrals.

Using this method, we obtain
\begin{equation}
\RE = -\frac{1}{2\beta^2},
\label{reponseharmonic}
\end{equation}
and
\begin{equation}
\mathrm{Var}(E) = \frac{1}{2\beta^2} 
+ \frac{f^2}{\beta (\kappa + 2\alpha \gamma)} 
+ \frac{\alpha \gamma f^4}{(3\kappa + 2\alpha \gamma)(\kappa + 2\alpha \gamma)^2}.
\end{equation}
To compute the Fisher information $\mathcal{I}_\beta$, we evaluate the integral in Eq.~\eqref{Px} and calculate $\mathcal{I}(\beta)$ numerically. 
In Fig.~\ref{Fig1}, the parameters are set to $\kappa = 1$, $\alpha = 0.4$, $\gamma = 1$, and $\beta = 1$.

\subsection{Calculations for the kinetic proofreading network}

The five states are labeled in the order $\mathrm{E}=1$, $\mathrm{ER}=2$, $\mathrm{ER}^* = 3$, $\mathrm{EW}=4$, and $\mathrm{EW}^* = 5$. 
The energies are given by $E_1 = E_2 = E_3 = 0$ and $E_4 = E_5 = \Delta E$.
The transition rates depend on the thermodynamic parameters $\beta$, $\Delta E$, and $\Delta \mu$, as well as on the kinetic parameters $k$, $\phi$, and $\gamma$. 
We use the same parametrization of rates as in Ref.~\cite{hart15}. The full stochastic generator reads
\begin{align}
\left(
\begin{smallmatrix}
-2k - 2\phi k  & k & \phi k & \phi k {\rm e}^{\beta \Delta E} & k {\rm e}^{\beta \Delta E} \\
k & -k - \gamma & \gamma {\rm e}^{-\beta \Delta \mu} & 0 & 0\\
\phi k & \gamma & -\phi k - \gamma {\rm e}^{-\beta \Delta \mu} & 0 & 0 \\
\phi k & 0 & 0 & -\phi k {\rm e}^{\beta \Delta E} - \gamma {\rm e}^{-\beta \Delta \mu} & \gamma \\
k & 0 & 0 & \gamma {\rm e}^{-\beta \Delta \mu} & -k {\rm e}^{\beta \Delta E} - \gamma
\end{smallmatrix}
\right).
\end{align}

The NESS distribution is obtained as the right eigenvector of this matrix associated with the zero eigenvalue. The full analytical expression is too cumbersome to display here. 
From the stationary distribution, all quantities entering the bound can be computed.
The average energy and its second moment are given by $\langle E \rangle = \Delta E \, (P_4 + P_5)$ and $\langle E^2 \rangle = \Delta E^2 \, (P_4 + P_5)$, respectively.
From these expressions, $\RE$ and $\mathrm{Var}(E)$ are obtained straightforwardly. The Fisher information $\Ib$ is computed as $\mathcal{I}_\beta = -\sum_{i=1}^5 P_i \, \partial_\beta^2 \ln P_i$ .
In the calculations shown in Fig.~\ref{Fig2}, the parameters are set to $k = \gamma = \beta = 1$, $\phi = 10^{-4}$, and $\Delta E = 5$.

\subsection{Calculations for the KMP model}

The KMP model \cite{kipn82} is defined as follows. The variable $i = -L, -L+1, \ldots, L$ labels the sites of a one-dimensional lattice 
of size $N = 2L+1$. The temperatures at the boundaries are $T_-$ (left) and $T_+$ (right). The energy at site $i$ is denoted by $e_i$. 
Two neighboring sites redistribute their total energy $\epsilon = e_i + e_{i+1}$ according to the rule $(e_i = \lambda \epsilon, \qquad e_{i+1} = (1 - \lambda) \epsilon)$,
where $\lambda$ is a random variable uniformly distributed in $[0,1]$. 
At the boundaries $i = -L$ and $i = L$, the energy is refreshed according to equilibrium distributions at temperatures $T_-$ and $T_+$, which is
$P(e_{\pm L}) = \frac{1}{T_\pm} \exp\left(-\frac{e_{\pm L}}{T_\pm}\right)$. In equilibrium, corresponding to $T_+=T_-$, the distribution of the full
energy profile is a simple product \cite{kipn82,bert05,bert15b}.

We define the empirical temperature distribution 
\begin{align}\label{def TN}
	T_N(x) \equiv \frac{1}{N} \sum_{i=-L}^{L} e_i \delta\left(x-\frac{L+i}{2L}\right),
\end{align}
where $x=0$ and $x=1$ correspond to $i=-L$ and $i=L$, respectively. Since we set $k_B=1$, temperature and energy have the same dimension. The field $T_N(x)$ is a fluctuating quantity, and the NESS is described by a probability functional over profiles.

The average temperature profile in a hydrodynamic limit is given by \cite{kipn82}
\begin{align}\label{T1980}
	\bar{T}(x) \equiv  \lim_{N\rightarrow\infty}\langle T_N(x) \rangle = T_- + x(T_+ - T_-).
\end{align}
Then the rescaled empirical temperature is defined as
\begin{align}\label{T tilde}
	\tilde{T}_N(x) \equiv \sqrt{N}\left(T_N(x)-\bar{T}(x)\right).
\end{align}

To compute fluctuations and the Fisher information, we use macroscopic fluctuation theory \cite{bert05}. In the large-$N$ limit, the probability functional has the large-deviation form
\begin{align}\label{Pasy}
P\left[\tilde{T}_N=\tilde{T}\right] \asymp \exp\left[-N S\left[\bar{T} + \frac{\tilde{T}}{\sqrt{N}}\right]\right],
\end{align}
where $S$ is the large-deviation functional. Although the full large-deviation functional $S$ is generally nonquadratic \cite{bert05}, its local expansion about the typical profile $\bar{T}$ is quadratic on the scale $T_N-\bar{T} \sim \mathcal{O}(N^{-1/2})$. For large $N$, we write for the leading order
\begin{align}\label{ST}
S\left[\bar{T} + \frac{\tilde{T}}{\sqrt{N}}\right] \approx \frac{1}{2N} \int \mathrm{d}x \mathrm{d}y \, \tilde{T}(x) C^{-1}(x,y) \tilde{T}(y),
\end{align}
where $C^{-1}(x,y)$ is the inverse covariance operator. Higher-order contributions vanish as $N \to \infty$. The covariance in the large $N$ limit is defined as
\begin{align}
	C(x,y) \equiv \lim_{N\rightarrow \infty} \left\langle \tilde{T}_N(x)\tilde{T}_N(y) \right\rangle,
\end{align}
and has the explicit form \cite{bert05}
\begin{align}\label{eqcxy}
C(x,y) = \bar{T}(x)^2 \delta(x-y) + (T_+ - T_-)^2 G(x,y),
\end{align}
where $G(x,y)$ is the Green's function \cite{bert05} satisfying $\partial_x^2 G(x,y) = -\delta(x-y)$ and $G(0,y) = G(1,y) = 0$. This function has explicit form 
\begin{align}\label{G}
	G(x,y) = \begin{cases} 
		x(1-y), & x <y \\
		y(1-x).   & x >y
	\end{cases}
\end{align}

We now compute the quantities appearing in Eq.~\eqref{eqmain}, using $\beta = 1/T_-$ and $\mathcal{F}_q = 1 - T_-/T_+$, with $T_+ \ge T_-$.
The total energy $E\equiv\sum_{i=-L}^{L} e_i$ is also given in terms of the empirical temperature $T_N(x)$,
\begin{align}
	E = N \int_{0}^{1} {\rm d}x T_N(x).
\end{align}
From the hydrodynamic limit \eqref{T1980}, one calculates the average
\begin{align}
	\lim_{N\rightarrow \infty} \frac{1}{N}\langle E\rangle = \int_{0}^{1} {\rm d}x \bar{T}(x) = \frac{1}{2\beta} \left(1+\frac{1}{1-\mathcal{F}_q}\right),
\end{align}
from which we obtain
\begin{align}\label{eqREKMP}
\frac{1}{N}\partial_\beta \langle E \rangle 
= -\frac{1}{2\beta^2}\left(1 + \frac{1}{1-\mathcal{F}_q}\right).
\end{align}

Integration of the rescaled empirical temperature \eqref{T tilde} yields
\begin{align}
	\int_{0}^{1} {\rm d}x \tilde{T}_N(x) = \frac{1}{\sqrt{N}}E - \sqrt{N}\frac{T_+ + T_-}{2},
\end{align}
from which one has at large $N$
\begin{align}
	E - \langle E\rangle \sim \sqrt{N} \int_{0}^{1} {\rm d}x \tilde{T}_N(x),
\end{align}
hence the variance of the total energy is written as
\begin{align}
\mathrm{Var}(E) = N \int \mathrm{d}x \mathrm{d}y \, C(x,y),
\end{align}
with $C(x,y)$ from Eq.~\eqref{eqcxy}. Using the explicit form of the Green's function \eqref{G},
we find
\begin{align}\label{eqFlucEKMP}
\mathrm{Var}(E) = \frac{N}{12\beta^2}
\left[ 5(1-\mathcal{F}_q)^{-2} + 2(1-\mathcal{F}_q)^{-1} + 5 \right].
\end{align}
This result contains local and nonlocal contributions, the latter arising from nonequilibrium correlations \cite{bert05}.

We now compute the Fisher information $\mathcal{I}_\beta$. This calculation, based on macroscopic fluctuation theory, is an original contribution. From Eqs.~\eqref{Pasy} and \eqref{ST}, we obtain
\begin{align}
\partial_\beta \ln P[T] \approx N \int \mathrm{d}x \mathrm{d}y \, \partial_\beta \bar{T}(x) \, C^{-1}(x,y) \, \delta T(y),
\end{align}
where we have used the symmetry of $C^{-1}(x,y)$ and neglected sub-leading terms in $N$ that are in the derivative of $C^{-1}(x,y)$ with respect to $\beta$ . Using Eq.~\eqref{eqdefI}, we find
\begin{align}
\mathcal{I}_\beta = N \int \mathrm{d}x \mathrm{d}y \, \partial_\beta \bar{T}(x)\, C^{-1}(x,y)\, \partial_\beta \bar{T}(y).
\end{align}
Using $\partial_\beta \bar{T}(x) = -\bar{T}(x)/\beta$, this formula becomes
\begin{align}\label{IbCinv}
\mathcal{I}_\beta = \frac{N}{\beta^2} \int \mathrm{d}x \mathrm{d}y \, \bar{T}(x) C^{-1}(x,y) \bar{T}(y).
\end{align}

To evaluate this expression, we avoid computing $C^{-1}$ explicitly. We define 
\begin{align}
	\Theta(x)\equiv \int_0^1 C^{-1}(x,y) \bar{T}(y)\,\mathrm{d}y,
\end{align}
which satisfies
\begin{align}\label{intC}
\int_0^1 C(x,y)\Theta(y)\,\mathrm{d}y = \bar{T}(x).
\end{align}
Let 
\begin{align}
	W(x)\equiv\int_0^1 G(x,y)\Theta(y)\,\mathrm{d}y,
\end{align}
with $W(0)=W(1)=0$. Then
\begin{align}
W''(x) = -\Theta(x).
\end{align}
Using Eqs.~\eqref{eqcxy} and \eqref{intC}, we obtain
\begin{align}
\bar{T}(x)^2 W''(x) - (T_+ - T_-)^2 W(x) = -\bar{T}(x).
\end{align}
Since $\bar{T}(x)$ is linear, we rewrite this equation in terms of $\bar{T}$:
\begin{align}\label{WODE}
\bar{T}^2 \frac{\mathrm{d}^2 W}{\mathrm{d}\bar{T}^2} - W = -\frac{\bar{T}}{(T_+ - T_-)^2}.
\end{align}
The general solution is
\begin{align}\label{Wsol}
W(\bar{T}) = A \bar{T}^{\lambda_+} + B \bar{T}^{\lambda_-} + \frac{\bar{T}}{(T_+ - T_-)^2},
\end{align}
where $\lambda_\pm\equiv \frac{1 \pm \sqrt{5}}{2}$ and
\begin{align}\label{A,B}
		& \mathcal{A} = \frac{T_- T_+^{\lambda_-} - T_-^{\lambda_-\lambda_-} T_+}{\Delta T^2 \left(T_-^{\lambda_-} T_+^{\lambda_+}-T_-^{\lambda_+} T_+^{\lambda_-}\right)},\nonumber\\ 
		 & \mathcal{B} = \frac{T_-^{\lambda_+} T_+ - T_- T_+^{\lambda_+}}{\Delta T^2 \left(T_-^{\lambda_-} T_+^{\lambda_+}-T_-^{\lambda_+} T_+^{\lambda_-}\right)}.
	\end{align}
Finally,  obtains $\Ib$ in Eq. \eqref{IbCinv} then becomes,
\begin{align}\label{eqIbKMP}
\mathcal{I}_\beta = \frac{N}{\beta^2}
\left[
-\frac{1}{2}
+ \frac{\sqrt{5}}{2}
\frac{\cosh(z/2)\cosh(\sqrt{5}z/2)-1}{\sinh(z/2)\sinh(\sqrt{5}z/2)}
\right],
\end{align}
where $z = \ln\left(\frac{T_+}{T_-}\right) = -\ln(1-\mathcal{F}_q)$.

The results shown in Fig.~\ref{Fig3} use Eqs.~\eqref{eqREKMP}, \eqref{eqFlucEKMP}, and \eqref{eqIbKMP} with $\beta=1$.

\acknowledgments
SL and ACB thank the financial support from the NSF through the grant  DMR-2424140. RC is supported buy the project RETENU ANR-20-CE40 of the French National Research Agency
(ANR).


\bibliographystyle{apsrev4-1}

\bibliography{refs}

\end{document}